\documentclass[12pt,preprint]{aastex}

\newcommand\lam{\mbox{$\:\lambda $ }}
\newcommand\lamlam{\mbox{$\:\lambda\lambda $ }}
\newcommand\ha{{H$\alpha$}}

\newcommand\kms{\:\rm{\,km\,s^{-1}}}

\newcommand\perpix{\:{\rm pixel}^{-1}}

\newcommand\eg{{\it e.g.}}

\newcommand\etal{et\thinspace al.}

\newcommand\feii{\ion{Fe}{2}}
\newcommand\feiii{\ion{Fe}{3}}

\newcommand{\SIii}{\ion{Si}{2}}
\newcommand{\SIiii}{\ion{Si}{3}}
\newcommand{\SIiv}{\ion{Si}{4}}

\slugcomment{Published in {\it The Astrophysical Journal},
\newline Vol. 624, pp 189--197, 2005 May 1}

\shorttitle{Probing Core Samples through SN~1006}
\shortauthors{Winkler, et al.}

\begin{document}


\title{Probing Multiple Sight Lines through the SN~1006 Remnant by UV
Absorption Spectroscopy}


\author{P. Frank Winkler\altaffilmark{1}}
\affil{Department of Physics, Middlebury College, Middlebury, VT 05753}
\email{winkler@middlebury.edu}

\author{Knox S. Long\altaffilmark{1}}
\affil{Space Telescope Science Institute, Baltimore MD 21218}
 \email{long@stsci.edu}

\author{Andrew J. S. Hamilton}
\affil{University of Colorado, Boulder CO 80309}
\email{andrew.hamilton@colorado.edu}

\author{Robert A. Fesen}
\affil{Dartmouth College, Hanover NH 03755}
\email{fesen@snr.dartmouth.edu}

\altaffiltext{1}{Visiting Astronomer, Cerro Tololo Inter-American Observatory.
CTIO is operated by AURA, Inc.\ under contract to the National Science
Foundation.}


\begin{abstract}

Absorption-line spectroscopy is an effective probe for cold ejecta
within an SNR, provided that suitable background UV sources can be
identified.  For  the SN~1006 remnant we have identified four such sources,
in addition to the much-studied Schweitzer-Middleditch (SM) star.  We have used STIS
on {\it HST} to obtain UV spectra of all four sources, to study ``core samples'' of the SN~1006 interior.
The line of sight closest to the center of the SNR shell, passing only 2\farcm 0 away,
is to a $V = 19.5$\  QSO at  $z = 1.026$\@.  Its spectrum shows broad
\feii\ absorption lines, asymmetric with red wings broader than blue.  The similarity of these
profiles to   those seen in the SM star,
which is 2\farcm8 from the center in the opposite direction, confirms
the existence of a bulge on the far side  of SN~1006.
The \feii\ equivalent widths in the QSO spectrum are $\sim 50\%$\ greater than
in the SM star, suggesting that somewhat more iron may be present within SN~1006
than studies  of the SM star alone have indicated, but this is still far
short of what most SNIa models require.

The absorption spectrum against a brighter $z = 0.337$\ QSO seen at 57\% of the shell radius
shows broad silicon absorption lines but no iron other than narrow, probably interstellar lines.
The cold iron expanding in this direction must be confined within
$v \lesssim 5200 \kms$, also consistent with a high-velocity bulge on the far side only.
The broad silicon lines indicate that the silicon layer has expanded beyond this point,
and that it has probably been heated by
a reverse shock---conclusions consistent with the clumpy X-ray structure and anomalous
abundances observed from {\it Chandra} in this region.
Finally, the spectra of
two $\sim{\rm  A0V}$\  stars near the southern shell rim show no broad or unusually
strong absorption lines,
suggesting that the low-ionization ejecta are confined within 83\% of the shell radius, at least at the
azimuths of these background sources.

\end{abstract}




\section{Introduction}

Young remnants of supernovae provide invaluable opportunities for probing
supernova ejecta and for making the connection between the supernovae and
their progenitor stars.  Spectroscopy in the optical, ultraviolet, and
X-ray regimes has achieved dramatic success at identifying emission lines
from virtually all the important chemical elements expected  in the
ejecta of either core-collapse or white-dwarf
supernovae.  Once the pyrotechnics
of the original supernova display have subsided, ejecta material that we
see radiating emission lines must first be excited---usually by a
strong reverse shock that propagates inward (in
the Lagrangian sense) through the expanding ejecta.  Ejecta material that
has {\it not} been shocked will remain cold and will not radiate, thus
leaving it invisible through emission spectroscopy no matter how exquisite
the instruments astronomers bring to bear.  Large quantities of cold
material likely reside inside many supernova remnants (SNRs), and this is
especially likely in remnants where the surrounding
interstellar medium (ISM) has extremely low density, so secondary shocks
are slow to develop.

An alternative to emission-line spectroscopy that can probe cold material
inside an SNR is {\it absorption-line} spectroscopy, but this requires the
presence of a suitable background light source that can be seen shining through the SNR
shell.  Since the strong permitted lines from low-ionization species lie
almost exclusively in the ultraviolet, ``suitable'' means UV-bright, with
the additional requirement of relatively low absorption along the line of
sight.  UV absorption spectroscopy has been exploited to probe a only handful of
large, nearby SNRs, \eg, Vela \citep{jenkins76, jenkins84, jenkins98},  the
Monoceros Loop \citep{welsh01}, and the Cygnus Loop \citep{blair04}.
These are quite different from SN~1006 in that all have old, highly evolved shells,
and the absorbing matter is mainly
interstellar material swept up and heated by the SNR shock.

More interesting from the standpoint of shedding light on supernovae and
nucleosynthesis are {\it young} remnants, but the relative rarity and small sizes
of these make absorption spectroscopy more difficult to employ.   So far one
young Galactic SNR---the SN~1006 remnant = G327.6+14.6---has been probed
through UV absorption along a single line of sight.   In addition, the SN~1885 remnant
in M31, discovered through its absorption in \ion{Fe}{1} \lam 3860
and \ion{Ca}{2}\lamlam 3934, 3968
by \citet{fesen89}, has also been studied spectroscopically in the UV \citep{fesen99}.
For SN~1006,  \citet{sch80} identified an OB subdwarf
star (henceforth the SM star), located only 2\farcm8  from the projected
center of the 15\arcmin\ radius
shell, at a distance that is probably only  slightly greater than the
$2.18\pm 0.08$\ kpc distance \citep{win03} to the remnant itself.

The opportunity to use the SM star to study SN~1006 ejecta in absorption
was first realized by \citet{wu83} using {\it IUE}\@.   This and
follow-up  {\it IUE} observations  \citep{fes88} revealed the presence of strong, broad
\feii\ absorption lines centered near zero velocity.  The {\it IUE}
observations also showed broad absorption lines in the far-UV, which
\citet{wu83} identified as redshifted \SIii, \SIiii, and \SIiv.
While clearly indicating the presence of cold iron within SN~1006, the inferred
mass fell well short of  the several tenths of a solar mass expected from
a Type Ia supernova, which the preponderance of present evidence indicates that
the SN~1006 event must have been
\citep{schaefer96,stephenson02,win03}.
\citet{blair96} used the {\it Hopkins Ultraviolet Telescope} to explore the
spectral region below Ly $\alpha$, where they found only very marginal evidence
for \feiii\ $\lambda\: 1123$\ absorption, indicating that the ``missing'' iron must,
if present, be mainly ionized more highly than Fe$^{++}$.
Subsequently, \citet{wu93, wu97} used the Faint Object Spectrograph (FOS)
on the {\it Hubble} Space Telescope to observe the SM star with significantly
higher signal to noise than had been possible from IUE, revealing weaker
absorption lines from the ejecta as well as narrow stellar and interstellar
absorption lines.   From a detailed analysis of these data, \citet[henceforth HFWCS]{hamilton97}
concluded that the profiles require a strong back-to-front asymmetry
in the ejecta and that both shocked and unshocked silicon are required.
They inferred a total silicon mass $M_{\rm Si} \approx 0.25\; M_{ \sun}$
and an iron mass $M_{\rm Fe} \approx 0.044\; M_{ \sun}$,
with a 3$\sigma$\ upper limit of  $M_{\rm Fe} \lesssim 0.16\; M_{ \sun}$.
Even the upper limit for the iron mass is significantly less than the 0.3 to 0.8 $M_{ \sun}$\ expected
from current models of Type Ia SNe \citep[\eg][]{hoeflich98, bravo04}.  Since the observations
were limited to a single line of sight through the
remnant, however, the mass estimates were based on a model that assumed
azimuthal symmetry about an axis along the line of sight through the center
of the shell.

Additional lines of sight---``core
samples''---through SN~1006 would have obvious value in elucidating the remnant structure.
\citet{win97a} identified a second UV ``light bulb'' behind SN~1006:  the
QSO 1504--4152, a $z=0.337$\ quasar with $V = 18.3$\ located 9\arcmin\ NE of
the remnant center.  In this paper we report the first UV spectra from
this object, as well as for three other background sources:  a fainter
($V \approx 19.5$) QSO only 2\farcm 1 north of the shell center, and two
early-type stars located just inside the southern limb of the remnant.

\begin{figure}[htbp]
\includegraphics[width=6.5in]{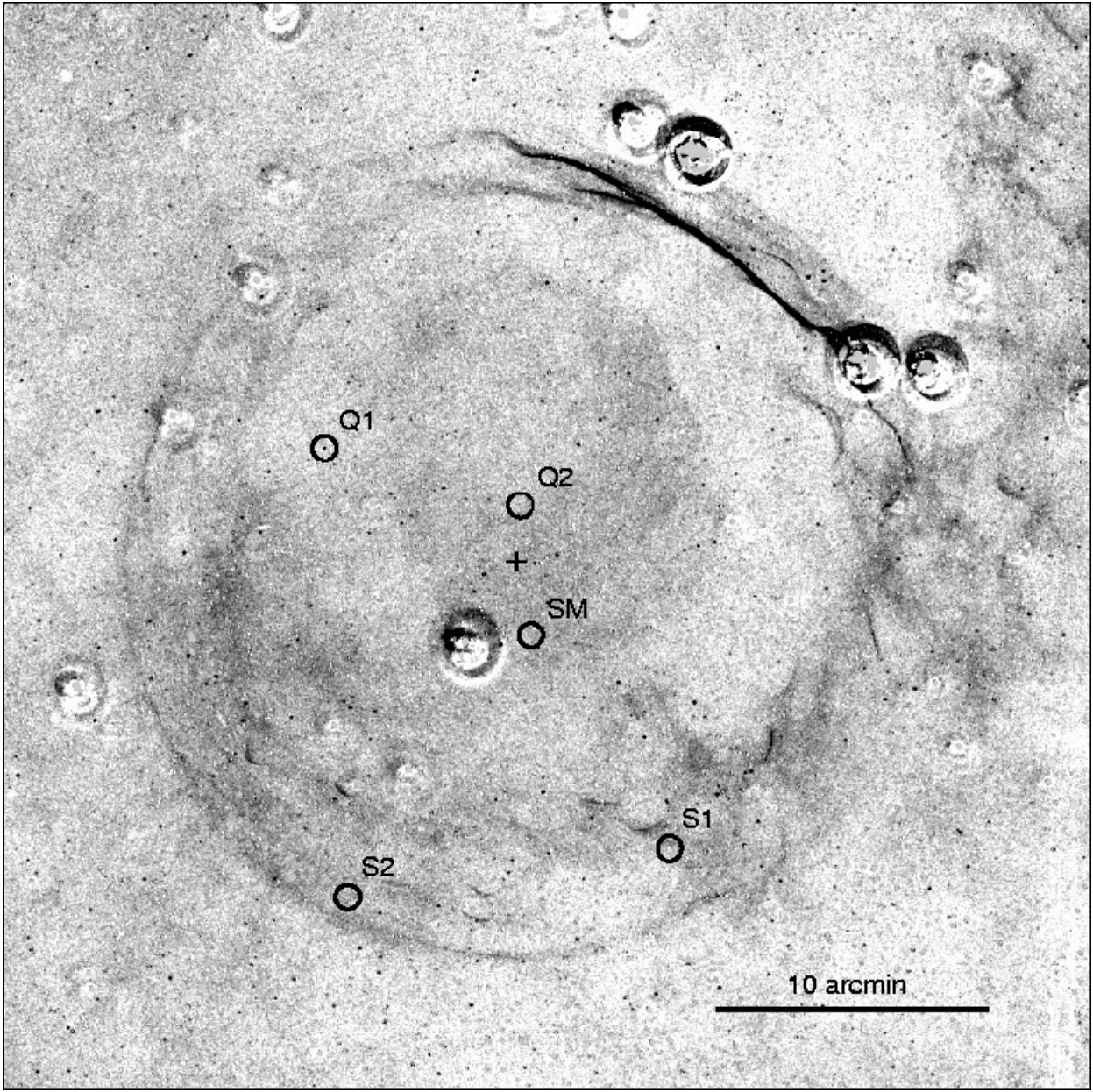}
\caption{
\label{fig1a}
The background UV sources are marked on this deep, continuum-subtracted
\ha\ image of SN~1006, which
shows the faint non-radiative filaments that delineate the primary
shock around most of the shell \citep{win03}.  All the
sources clearly lie within the projected remnant.
North is up, and east to the left.  The cross indicates the geometric center
of the optical shell.
}
\end{figure}

\begin{figure}[htbp]
\figurenum{1b}
\includegraphics[width=6.5in]{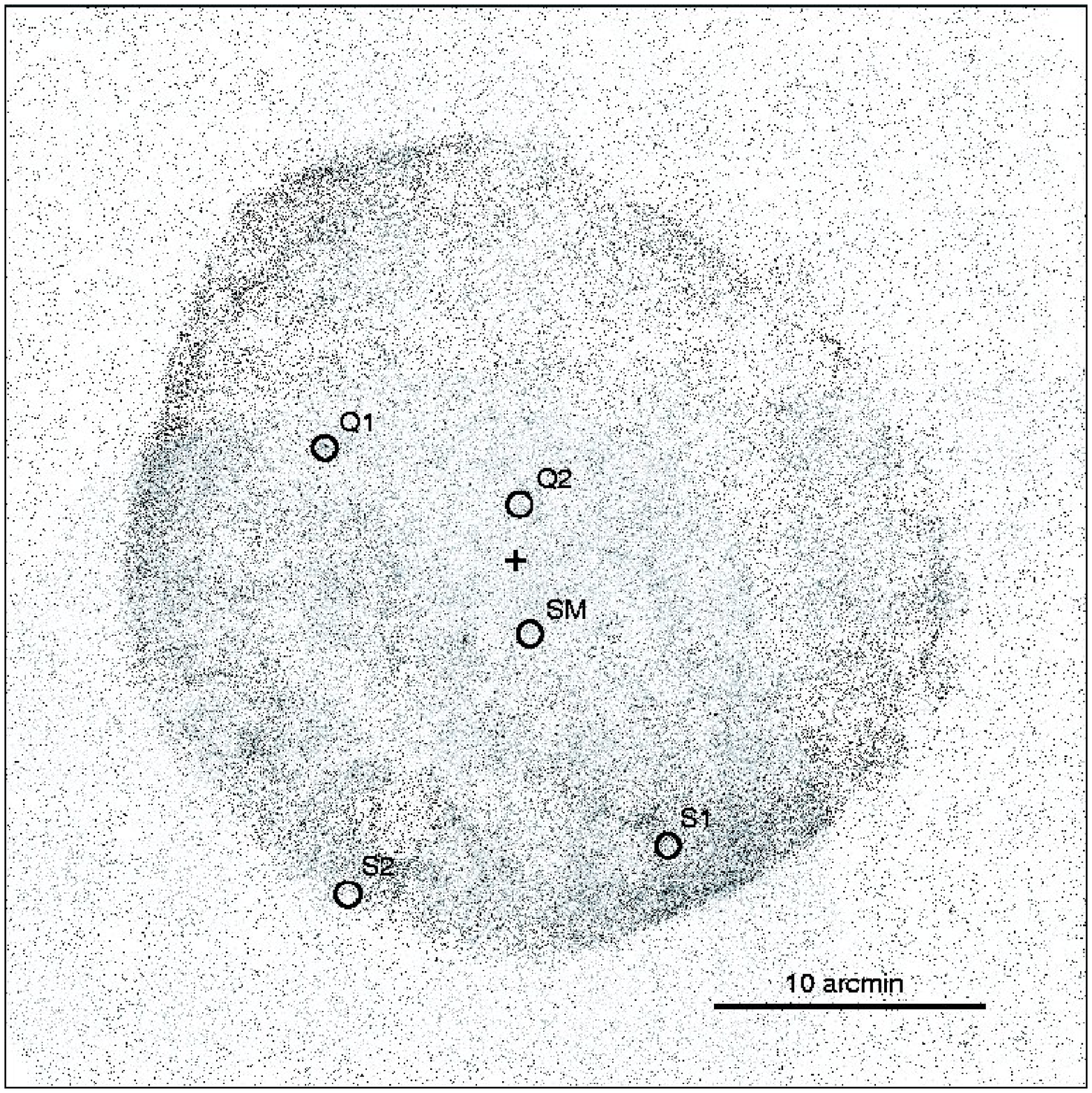}
\caption{
\label{fig1b}
{\it ROSAT}\ \ HRI X-ray image of SN~1006 \citep{win97a}.  The circles mark the locations
of background UV sources (Table 1), and the cross marks the remnant center, as determined
from the \ha\ shell \citep{win03}.
}
\end{figure}

\section{Observations}

\subsection{Optical Identification of Background Sources}

The QSO 1504--4152, hereafter Q1, first drew our attention as a
faint, unresolved X-ray source in a {\it ROSAT} HRI observation of
SN~1006, and we subsequently  identified it as a quasar at
$z=0.337$\ in spectra obtained at the CTIO 1.5-m telescope \citep{win97b, win97a}.
Buoyed by this success, we then attempted to identify other
background UV sources. In 1997 February we obtained a set of {\it
UBV} images obtained of the SN~1006 region  from the
CTIO 0.9-m telescope. Our four contiguous 14\arcmin\ square fields
covered  almost all of the 30\arcmin\ diameter
shell the SNR\@. From these images we identified 7 new candidates
that had relatively blue colors ($U-B\ {\rm and}\ B-V$) and that were bright
enough for spectroscopic observation at
UV wavelengths from  {\it HST} to be practicable.

We then obtained spectra of all 7 optical candidate, plus the SM
star and QSO 1504--4152, with the 4-m Blanco telescope at CTIO in
good observing conditions in 1998 June 24--26 (UT).  We used the
RC spectrograph with the 527 line mm$^{-1}$\ KPGL3 grating, Blue
Air Schmidt camera, and the Loral 3K CCD to cover the range
3200--6700 \AA\ at a dispersion of $1.21\; {\rm\AA}\: {\rm
pixel}^{-1}$\@. With a slit width of 1\farcs 5, the resolution
(judged from arc lamp spectra and from night sky lines) was
3.9 \AA\@. We reduced the spectra using standard
IRAF\footnote{IRAF is distributed by the National Optical
Astronomy Observatories, which is operated by the  AURA, Inc.
under cooperative agreement with the National Science Foundation.}
procedures; we used exposures of several spectrophotometric
standard stars from \citet{hamuy92} for flux calibration.

The spectra allowed us to identify three additional objects that appeared to
be good candidates for ``UV lightbulbs" at distances beyond SN1006, and to
exclude the remainder.   Figures~\ref{fig1a} and \ref{fig1b} illustrate (in both \ha\ and X-ray images)
the locations of these objects, as well as those of
the previously identified QSO (Q1) and the SM star, relative to the SN~1006 shell.
One of the candidates,
hereafter Q2, is a $V = 19.5$, $z = 1.026$\ QSO, based on
the broad emission features centered at 5665 \AA\ and $ \sim 3860 $\  \AA,
which represent redshifted \ion{Mg}{2} $\lambda\: 2799$\ and
\ion{C}{3}] $\lambda\: 1909$, respectively.\footnote{The best-fit redshift
for the optical spectrum alone is $z = 1.024$, but  for the
UV-optical combination the best fit is $z = 1.026$\@.  We adopt the latter value
throughout this paper.}

\begin{figure}[htbp]
\begin{center}
\includegraphics[width=3.5in]{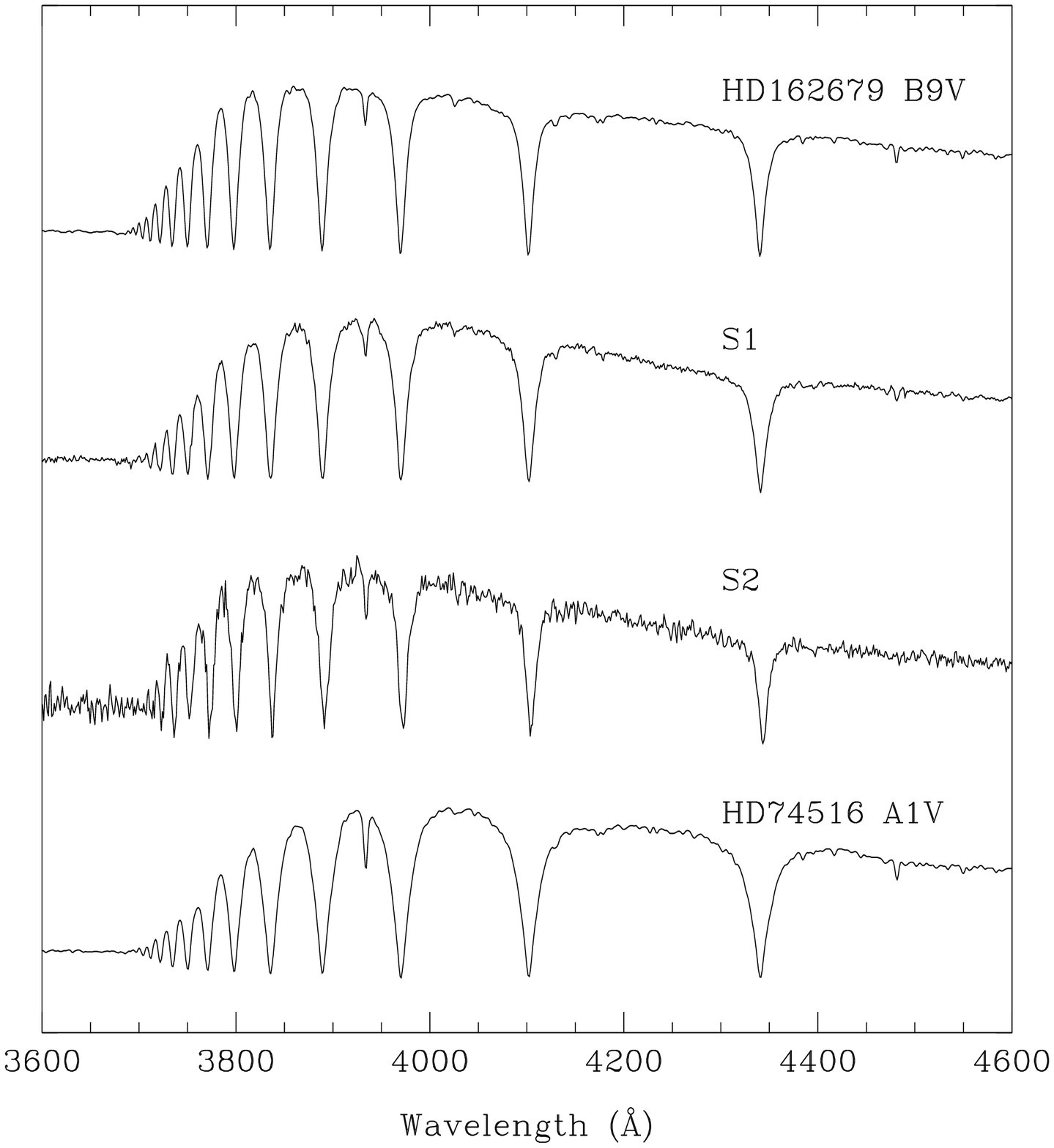}
\end{center}
\caption{
\label{fig2}
Optical spectra of the two stars, S1 and S2, that we have identified as background sources to
SN~1006.  For  comparison we show spectra from stars of slightly later and of comparable to slightly
earlier spectral types, both taken from the ESO Library of High-Resolution Spectra
\citep{bagnulo03} and smoothed to a resolution to match the resolution of our spectra.  The
strengths of the \ion{He}{1}\ $\lambda\: 4026$\ and \ion{Ca}{2}\ $\lambda\: 3934$, along
with the Balmer profiles,  indicate that both S1 and S2 are very close to A0V stars.}
\end{figure}

Two other candidates, hereafter designated S1 and S2, are stars located near the
rim of the SNR, both with effective temperatures of about 10,000 K
based on comparisons with stellar model spectra. The
Balmer line profiles and the strengths of weaker lines  indicate that
both are probably B9V to AOV stars, as shown in Figure~\ref{fig2}.
Similar comparisons indicate that the remaining candidates are nearby
DA white dwarfs, and hence uninteresting from the perspective of
using them for observing absorption in SN1006\@. Table 1 provides
positions and observational details for the three new objects, as well
as for the SM star and QSO 1504--4152.

Both QSOs are obviously background sources to SN~1006, and we believe
that both the stars are as well. Several lines of argument all
lead to a distance to SN~1006 near 2 kpc  \citep{schaefer96}.
\citet{win03}
used a combination of proper-motion measurements and the shock
velocity \citep{gha02} to obtain a geometric distance to
the remnant of 2.18$\pm$0.08 kpc, corresponding to a distance modulus
$(m - M)_0 = 11.69 \pm 0.08$\@.
Stars S1 and S2 have apparent magnitudes $V=13.54$\ and
16.61, respectively; if we assume that both are A0V stars with absolute
magnitude $M_V \approx 0.65$\ \citep{drilling00}
and extinction $A_V \approx 0.35$\ (comparable to that
for the SM star\footnote{Most estimates of  $A_V$\ to the SM star cluster
near this value:
0.32 \citep{sch80}, 0.31 \citep{wu93}, 0.34 \citep{blair96}, and 0.36 (HFWCS).
The one discrepant estimate is 0.48 by \citet{burleigh00}.}), then the two stars
have distance moduli 12.54 and 15.61, respectively, both comfortably beyond
SN~1006.   The exact spectral type does not change this
conclusion.  Even if S1, the brighter of the two stars,
were as late as A2V, with an effective temperature almost 1000 K
below that of our best-fit models, it would still be at a greater  distance than SN~1006.

(Note that the distance to the SM star itself is quite uncertain.  \citet {sch80}
estimated $(m - M)_0 = 10.2 \pm 1.8$, a distance range of 0.5 to 2.5 kpc.
\citet{burleigh00} later argued for a somewhat tighter range:  $d = 1.05\ {\rm to}
\ 2.1$\ kpc\@.  But unless the absorption lines in the SM star spectrum stem from
some pathological origin it must surely be beyond SN~1006.)

\subsection{STIS Spectra of the Background Sources}

Once a set of UV sources behind SN1006 had been identified, the next step
was to obtain spectra with {\it HST}\@.  Q1 had never been observed
previously with {\it HST}, and so we obtained spectra of it as
well as the three new candidates, Q2, S1 and S2. Our observations
with {\it HST} were carried out with STIS using the low-resolution
G140 or G230L gratings, the 52$\arcsec \times 0\farcs2$\
aperture, and the FUV or NUV MAMA detector operating in
photon-counting mode. The G140L spectra cover the wavelength range
1150-1740 \AA\ with a dispersion of $0.6\: {\rm \AA \perpix}$,
while the G230L spectra cover the range 1570-3180 \AA\ with a
dispersion of $1.6\;{\rm \AA} \perpix$\@. Given the velocity
widths of order $5000 \kms $\ in the ejecta of SN1006, lines from
the ejecta were expected to be (and are) well-resolved. All of the
sources, except for Q2 which is quite faint in the FUV, were
observed with both instrumental setups in order to obtain as
complete a description of the line-of-sight abundances as
possible. A log of the observations is presented in Table 2. There
were no anomalies during the observations themselves.

All of the data were processed using the standard STIS
pipeline and calibration files available in 2002 March in order to
extract time-averaged spectra.  As discussed below, both quasar
spectra show clear evidence of broad absorption lines at the
positions of transitions which had been detected in absorption in
the {\it IUE} and {\it HST} spectra of the SM star \citep{wu83, fes88, wu93, wu97}.
Neither of the two stars observed shows similar absorption features
in its spectrum.   If our assumption that the stars lie beyond SN~1006
is correct, this indicates that low-ionization ejecta are much less prominent along the lines of sight to
these objects, located near the periphery of SN~1006 shell, than along
the lines of sight that pass more centrally through the remnant.

\begin{figure}[htbp]
\begin{center}
\includegraphics[width=3.5in]{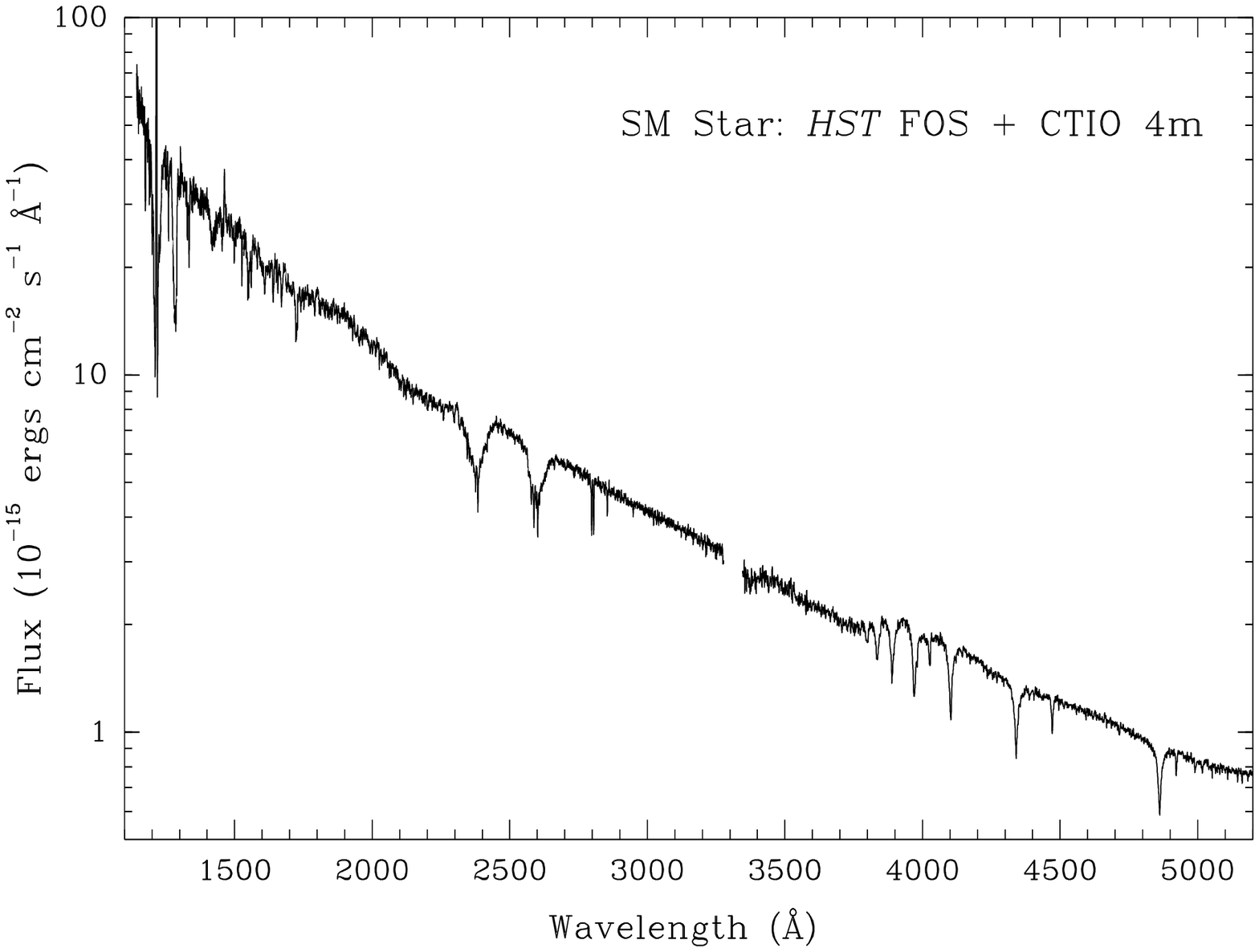}
\end{center}
\caption{
\label{fig3}
Combined UV-optical spectrum of the SM star.  The UV data represent a combination
of all the FOS data \citep{wu93,wu97}, taken with 3 different gratings, and the optical
data are from the CTIO 4m.   The absolute flux scales for the four individual spectra
have been adjusted relative to one another by up to 15\% in order to achieve matching
in the regions of overlap.
}
\end{figure}

\section{Results}

Since the benchmark for absorption in SN~1006 is the SM star, we show in Figure~\ref{fig3}
its complete UV-optical spectrum.  We combined the (previously published)
FOS data taken with three
gratings---G270H, G190H \citep{wu93} and G130H \citep{wu97}---with our new
data from the CTIO 4m.   Minor inconsistencies in the absolute flux levels from the
four separate observations were removed by scaling the levels to achieve consistency
in the regions of overlap.   We used scaling factors as follows:  G130H:  0.96, G190H: 0.89,
G270H: 1.00, 4m: 1.15.
As can be inferred from the necessity for such scaling,
there are systematic uncertainties as large as 15\% in the absolute flux shown in Figure~\ref{fig3}.

\begin{figure}[htbp]
\begin{center}
\includegraphics[width=3.5in]{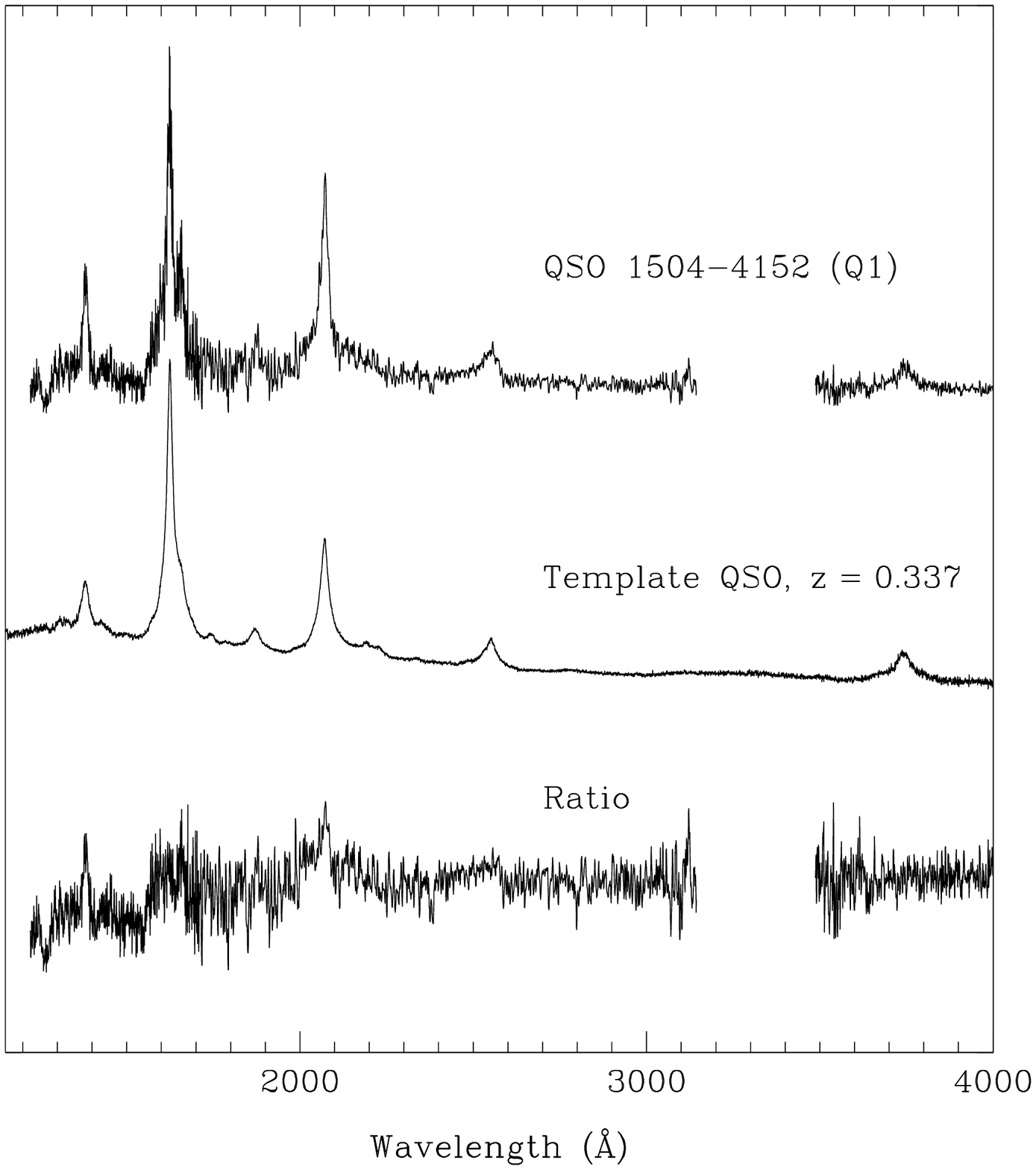}
\end{center}
\caption{
\label{fig4}
Comparison of the combined UV-optical spectrum from Q1 (top) with a template
QSO spectrum \citep{zheng97}, redshifted by $z=0.337$\ (center).  The bottom curve
is the Q1 spectrum, normalized by simply taking the ratio to the template.  The broad absorption
line near 1260 \AA, which we attribute to \SIii, is shown in detail in Fig.~\protect\ref{fig7}.}
\end{figure}

\begin{figure}[htbp]
\begin{center}
\includegraphics[width=3.5in]{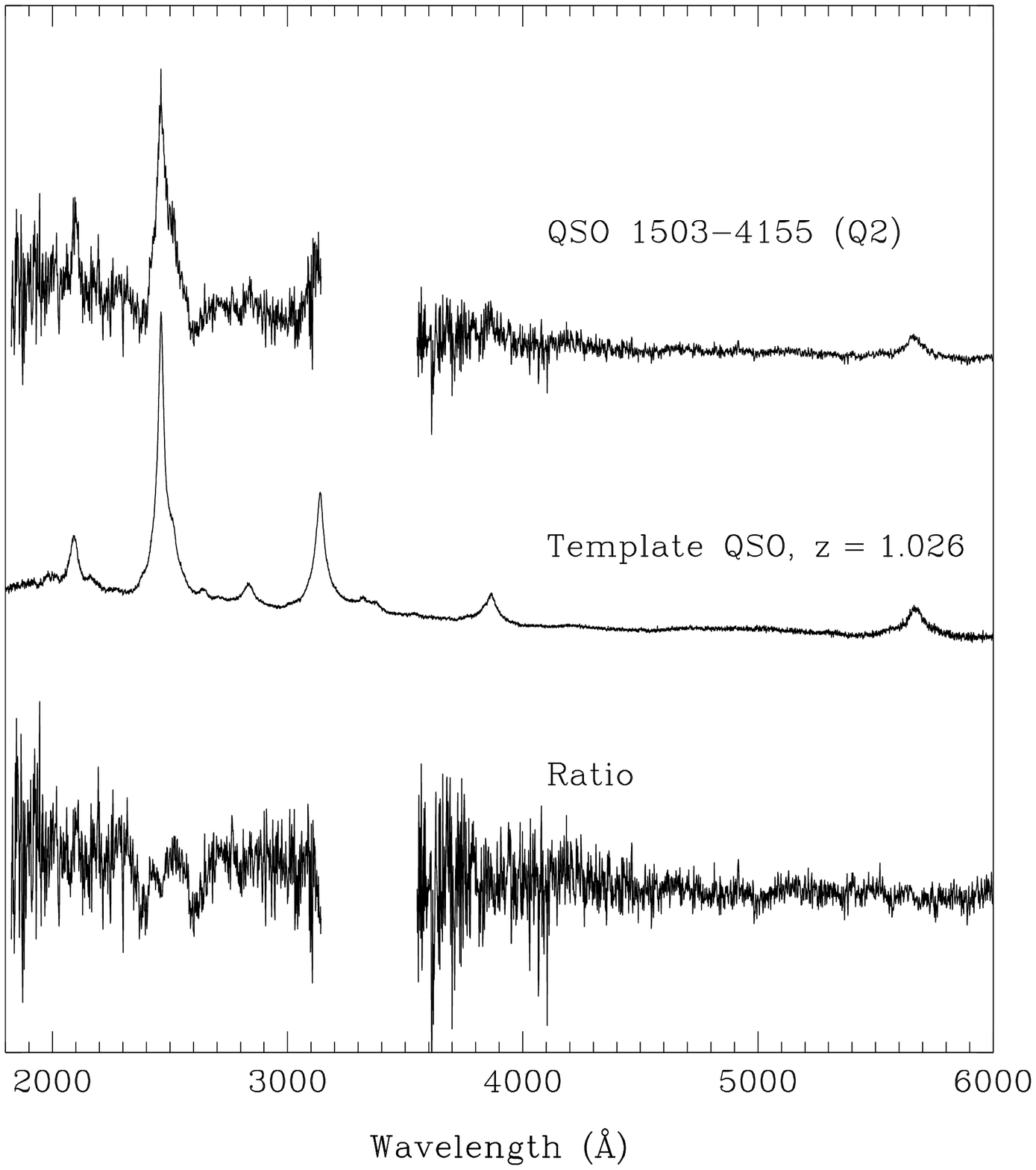}
\end{center}
\caption{
\label{fig5}
Comparison of the combined UV-optical spectrum from Q2 with the template QSO spectrum,  \citep{zheng97}, redshifted by $z=1.026$, as in Fig.~\protect\ref{fig4}.  Broad \feii\  absorption lines near 2400 \AA\
and 2600 \AA\ are evident.   Fig.~\protect\ref{fig6} shows a detailed view of this region.}
\end{figure}

In order to investigate any posited foreground absorption against the new sources,
we must compare the spectrum of each with a template (unabsorbed) spectrum from a similar source.
For the two QSO's, we have used as a template the composite QSO spectrum from
\citet{zheng97}, redshifted to match our background objects.   Figures~\ref{fig4} and \ref{fig5} show
the combinations of our 4m optical spectra and STIS UV spectra  for QSOs Q1 and Q2,
respectively.   We have dereddened these spectra by arbitrarily taking
$A_V = 0.35$, essentially the same as that for the SM star, under the assumption that the great
majority of the reddening/extinction occurs in the portion of the Galaxy  between us and
SN~1006, which lies at a distance of 550 pc above the Galactic plane.  We used the empirical
extinction function of \citet{cardelli89} for the dereddening.    Also shown in the figures
is the template QSO, redshifted to z = 0.337 and 1.026 for Q1 and Q2, respectively, and
normalized spectra obtained by simply dividing our combined spectra by the appropriate
template.  The match between our spectra and the template is quite good in both
cases---perhaps surprisingly so for combined spectra from different instruments on the
ground and in space.

\begin{figure}[htbp]
\begin{center}
\includegraphics[width=3.5in]{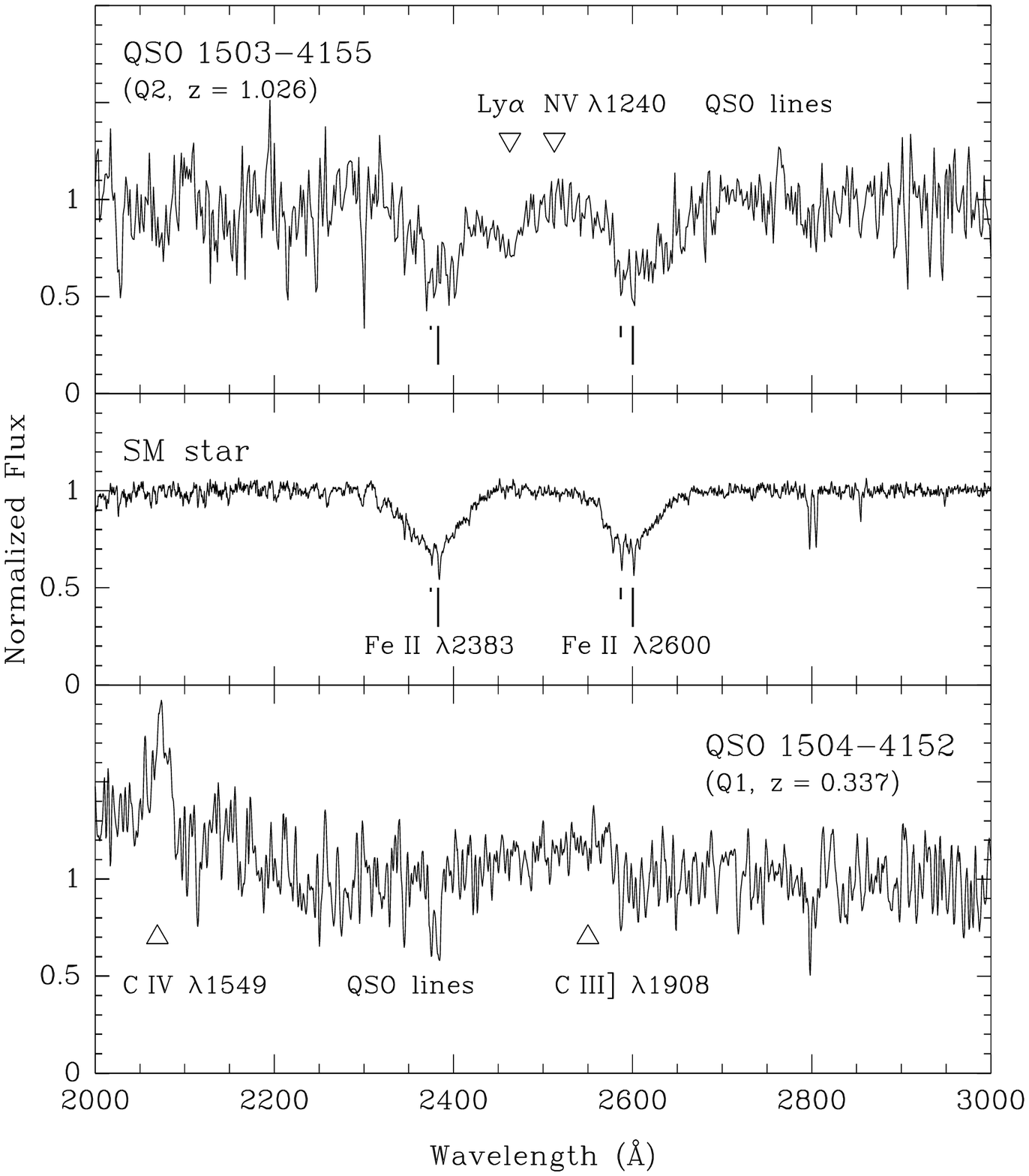}
\end{center}
\caption{
\label{fig6}
The normalized spectrum of Q2 in the region of prominent \feii\ lines (top) shows clear
evidence for broad  absorption features, with a relatively sharp blue edge
at $\sim -3000 \kms$, and a much more gradual red edge extending at least to
$8000 \kms$.ÊÊ (The line at 2600 \AA\  is much cleaner than the one at 2383 \AA,
since the latter is marred by near coincidence with redshifted Ly
$\alpha$\ and \ion{N}{5}\ emission from the QSO.)  The profile is similar to that
seen in the SM star spectrum (middle), also normalized to the continuum.
The Q1 spectrum (bottom) shows no evidence for broad \feii\  lines; the narrow lines
are probably of interstellar origin.
}
\end{figure}


\begin{figure}[htbp]
\begin{center}
\includegraphics[width=3.5in]{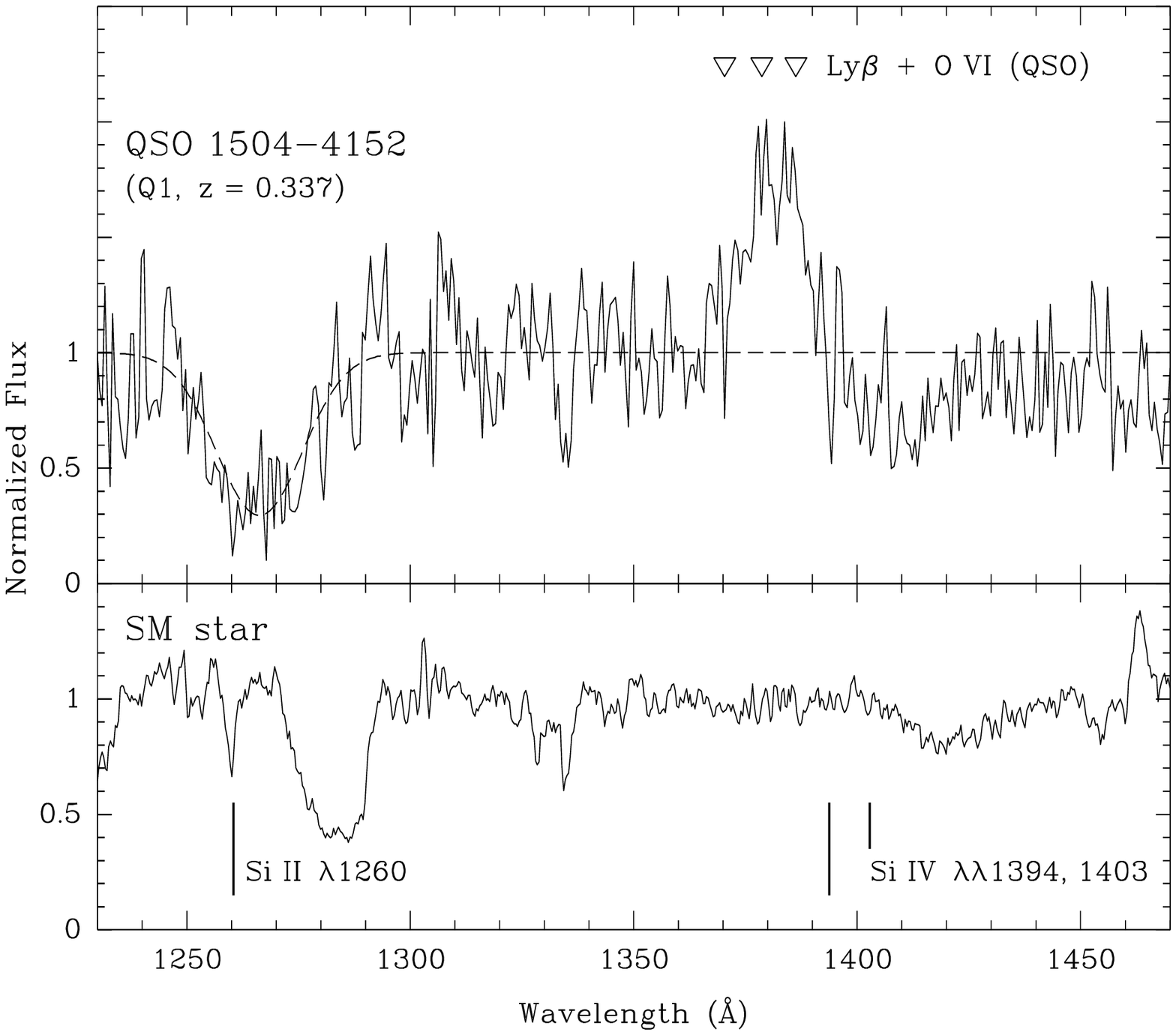}
\end{center}
\caption{
\label{fig7}
The normalized spectrum of Q1 in the region of \SIii\  and \SIiv\  lines.   The data
suggest broad  \SIii\  and possibly  \SIiv\   absorption.Ê As in the SM star spectrum,
absorption is almost all on the red-shifted side of the line.Ê Unfortunately, redshifted
Ly$\beta$\ + \ion{O}{6}\  emission confuse the absorption from
\SIiv \lamlam 1394, 1403.
}
\end{figure}

The clearest evidence for foreground absorption is seen in Q2, where broad
 absorption lines from \feii\ $\lambda\: 2383$\ and $\lambda\: 2600$\ are apparent,
 as shown in a plot of this region of the normalized spectrum in
 Figure~\ref{fig6}.\footnote{In actuality each of these two lines is the strongest component
 of a multiplet; see HFWCS for more details.}
 Also shown in Figure~\ref{fig6} is the {\it HST}-FOS spectrum from the SM star, normalized
 by a fit to the continuum based on featureless regions of the spectrum.  (This is
 very similar to that illustrated in Figures~\ref{fig4} and \ref{fig5} of HFWCS.)
  In the Q2 spectrum the line profile for the 2383 \AA\ line is confused by the proximity of
 redshifted Ly $\alpha$, which results in imperfect normalization of the spectrum near
 2460 \AA; the 2600 \AA\ line is considerably cleaner so we use that for most of our subsequent
  analysis.   The absorption
 profiles along the  lines of sight to the SM star and Q2 appear quite similar; the
  lines in both objects  show a relatively sharp blue-shifted edge and a more
  gradual slope on the red-shifted side.  Through detailed fits to the SM star spectrum,
HFWCS concluded that absorbing Fe$^+$\ is present at velocities from
  $-4200\ {\rm to\ } +7100 \kms$.
  The blue-shifted edge in the Q2 spectrum occurs at somewhat lower velocity:  for the
2600 \AA\  line, the edge occurs at $-2450 \pm 180 \kms$, while in the noisier
2383 \AA\  line, the blue edge is at $-2100 \pm 600 \kms$\@.

  The \feii\ absorption appears to be somewhat deeper in the Q2 spectrum than in
  the SM star.   The total equivalent width of the \feii\ line complex near 2600 \AA\ line in the Q2
  spectrum is $31\pm 6\:\rm\AA$.   This includes several narrow interstellar lines of  \feii\ and
  \ion{Mn}{2}\ in addition to the broadened lines at 2600 \AA\  and 2586 \AA\  from SN~1006 itself.
We removed the narrow lines and assumed that the broad lines are in the ratio of their
oscillator strengths \citep{morton03} to obtain an equivalent width of $23\pm 4\:\rm \AA$\ for
\feii \lam 2600, compared with 14.8 \AA\ for the same line in the SM star \citep{wu93}.

Also shown in Figure~\ref{fig6} is the normalized spectrum from Q1, which
shows no \feii\ absorption lines even approaching the width or strength of those
in the Q2 or SM star spectrum.  There is a pair of  narrow lines at wavelengths corresponding
to the two strongest members of the \feii  \lam 2383 triplet at essentially zero velocity,
but we interpret these as being primarily of interstellar origin.
While we cannot rule out a small amount of \feii\ associated with SN~1006, there is clearly
far less than that seen in the core samples toward the two more centrally located
background objects.

Figure~\ref{fig7} is a comparison of the region containing \SIii\  and \SIiv\   lines in Q1 and
the SM star.  Again the QSO spectrum is confused by the presence of redshifted lines in
Q1 itself, in this case Ly $\beta$\ + \ion{O}{6}, as well as Galactic Ly $\alpha$.
The cleanest and strongest line
is \SIii\ $\lambda\: 1260$, which is also the strongest absorption line in the SM star spectrum.
For the SM star, the salient features are that all of the absorption is redshifted:  there is  a
narrow component with a sharp
red edge at 1290 \AA, corresponding to $+7026\pm 10 \kms$,
and a broader component that
can be well fit by a Gaussian profile centered at $+5150 \pm 50
\kms$\ with a dispersion ($\sigma$) of $1140\pm 50 \kms$; these two components
may be attributed to unshocked and shocked \SIii, respectively
(Hamilton \etal, in preparation).   We find that the corresponding feature
in Q1 is comparable in equivalent width to that in the SM star spectrum.  The Q1 absorption
centroid is at $+1570 \kms$, significantly less than for the SM star, and the profile appears
to be more symmetrical and possibly a bit broader.  The line is consistent with
a Gaussian with FWHM 22.7 \AA, or dispersion $\sigma = 2300 \kms$, as
shown in the fit in Figure~\ref{fig7}.

Q2 is so faint that we did not attempt a spectrum with the G140L setup
 and thus have no  measurement of \SIii\  absorption along this
second line of sight  at roughly the same distance from
the SNR center as the SM star.  We discuss the possible implications
of both iron and silicon absorption for SN~1006 in the next section.

\section{Discussion}

The observation of similar absorption profiles in the spectra of Q2 and the SM star rules out
the possibility that the SM star might have been the donor star in the pre-SNIa
system that gave rise to SN~1006---an idea advanced by \citet{wellstein99} and discussed
by \citet{burleigh00}.   Since it is surely impossible for a donor star to have
a runaway velocity $\gtrsim 7000 \kms$, sufficient for it to have outrun the
expanding ejecta, this idea would require the SM star to be within the remnant shell,
with some exotic origin for the broad absorption lines in its spectrum.   As \citet{burleigh00}
themselves argued, detection of similar absorption lines in another background object
would exclude this possibility.  Q2 is just such an object.

\subsection{Iron Lines}

The lines of sight to Q2 and the SM star probe distinct
core samples through SN~1006 that are separated
by  4\farcm8 and located 2\farcm0 and 2\farcm8, respectively,  from the projected
center of the remnant.  The qualitative similarity in the iron absorption profiles against these
two background objects (Figure~\ref{fig6}) strongly suggests that the distribution of absorbing
material is similar and that small regional concentrations or partial shells of
iron-rich ejecta are not  responsible for the
observed profiles.   The asymmetric profiles for both sources
are consistent with the interpretation advanced
by HFWCS  that most of the \feii\  is unshocked ejecta
filling the region defined by the reverse shock, but that there is a
strong front-back asymmetry with the far side of the SNR expanding more
rapidly than the near side.
HFWCS further argued that the sharp blue edge for the 2600 \AA\
absorption against the SM star represents the free expansion radius of \feii\ that is just reaching
the reverse shock.   The observation of sharp blue edges for both the 2600 \AA\ and the
2383 \AA\ lines in the Q2 spectrum, at a somewhat different velocity than in the SM star
($- 2400 \kms$\ for Q2, $- 4200 \kms$\ for the SM star), strengthens this argument.  The
different velocities mean that the reverse shock has probably penetrated a bit farther
into the expanding iron on the near side of the shell in the direction toward Q2, probably
reflecting inhomogeneities in the ambient density of the ISM on the near side.

The apparent absence of \feii\ aborption  (other than narrow interstellar lines) against Q1,
which probes a core sample
8\farcm3 away from the remnant center, indicates that little if any of the iron
launched by the supernova explosion has advanced this far out.  At a distance
of 2.18 kpc, 8\farcm3 corresponds to a radius of 5.6 pc and an
average velocity of $5160\kms$\ over the 995 years from A.D. 1006 until the date of
our observations.  We expect that cold ejecta will have remained essentially
undecelerated since the explosion (otherwise they would have encountered
a shock and been heated), so the present velocity should be close to the average one.
If the iron ejecta are moving outward at $\lesssim 4200 \kms$, the velocity for the cold iron
on the {\it near} side of the remnant as inferred from the two central core samples,
then none would have reached the line of sight to Q1, consistent with our
observations.  This lends credence to  a model
calling for a high-velocity bulge on the far side of SN~1006.

HFWCS concluded, based on their detailed analysis of absorption along the
single line of sight to the SM star spectrum, that there is about $0.029\: M_\sun$\ of
\feii\ within SN~1006.   They argued that most of the iron should be unshocked and cold, and
estimate that the ratio ${\rm Fe/ Fe\:II} \simeq 1.5$\ to obtain a estimate for the total iron mass
$M_{\rm Fe} \simeq 0.044\: M_\sun$\ with a $3\:\sigma$\ upper limit of $M_{Fe} < 0.16\: M_\sun$.
Observations from HUT of low \feiii\  \lam 1123 absorption
in the SM star spectrum are also consistent with low ionization and a low total
iron mass \citep{blair96}.
Our observation of absorption against Q2 with similar profile but with equivalent width
$\sim 50\%$\ greater than  that against the SM star leads us to revise estimates
for the iron mass upward somewhat, to $M_{Fe} \simeq 0.06\: M_\sun$.  The second
line of sight removes some of the uncertainty associated with the models, so we
retain the same $3\sigma$\ upper limit of $M_{Fe} < 0.16\: M_\sun$.
The  fundamental
conclusion reached by all previous absorption studies of SN~1006,  that the iron mass
is much less than the several tenths of a solar mass  expected to be associated
with a Type Ia supernova remnant \citep[\eg][]{hoeflich98,  bravo04}, remains unchanged.
Indeed, the conclusion that iron is unexpectedly
scarce  in SN~1006  is solidified by demonstrating
that the meager amount of iron along the path to the SM star is not an anomaly.

Other possible support for a smaller amount of iron in the centers of some  SNeIa
than standard Type Ia models predict comes from a new analysis of
late-time infrared emission from SN 2003du,
which shows a flat-topped [\feii]  1.644 $\mu$m emission profile indicative
of a thick but hollow-centered iron ejecta shell (H{\" o}flich, et al., in preparation).
Less than the ``standard'' amount of iron in the center of the young SNIa remnant
0509-67.5  in the LMC might also
help explain its observed X-ray emission properties \citep{warren04}.
Finally, there have been theoretical suggestions  by \citet{kasen04}  that a
substantial hole in the ejecta might result when a white dwarf explodes
and engulfs its  nondegenerate companion star.
In any case, the UV absorption observations presented here for several
sight-lines through SN~1006 make a strong case by themselves that there is
significantly less iron than standard models predict in this almost certain SNIa remnant.

\subsection{Silicon Lines}

 In the SM star, \SIii\ \lam 1260 shows a highly redshifted profile.  Indeed it is this
profile, more than the \feii\  profiles, that requires a very asymmetric
distribution of material  along a line of sight passing near the center of the
SNR\@.   HFWCS argued that the profile comprises two
components:  the more redshifted component consists of material
still freely expanding inside the reverse shock.   The second component
represents post-shock ejecta, which are physically
located outside of the reverse shock  but have a lower velocity, since
 they have been decelerated in passing through that shock.  The absence of
 appreciable  \SIii\  absorption near zero velocity implies that there is little  silicon in much of
the region occupied by iron in the SNR.  Likewise, the spectrum of the SM star
also shows no evidence for  shocked or unshocked \SIii\  from the near side of the
SNR.  The possibility that some high velocity blueshifted \SIii \lam 1260 is hidden
in the red wing of \SIiii \lam 1206 is excluded by the absence of corresponding
blueshifted \SIii \lam 1527.


The strong \SIii\  absorption along the line of
sight to Q1, which passes 8\farcm3 from the center of the shell, indicates that much
of the expanding silicon has advanced farther out than this line of sight.
The observation that the \SIii\  absorption is redshifted in the spectra of both the
SM star and Q1  means
that the asymmetry observed in the \SIii\  absorption along the line of sight
to the SM star extends over much of the far side of the SNR, consistent with
a picture in which the ambient density on far side is lower than on the near side.
The shift in the  line centroid of only $1570 \kms $\ for Q1, rather than
$\sim  5000 \kms $\ as for the SM star,  can be understood qualitatively as a
projection effect:   the bulk velocity of the  material should be radial with
respect to the site of the SN explosion, so material we see farther from the
center must have a greater tangential component and a lesser one along
the line of sight.

The approximately Gaussian profile of
redshifted  \SIii\   absorption suggests that the line of sight to Q1
is almost tangent to the reverse shock.
If the line of sight to Q1 were well inside the reverse shock,
then the \SIii\ 1260\,\AA\ absorption profile
would be shaped more like a top hat, with sharp blue and red edges,
contrary to our observations.
Conversely,
if the line of sight from Q1 were well outside the reverse shock,
then the Si would have been collisionally ionized to high ionization stages
such as are observed in X-rays \citep{dyer01, long03},
and there would be no \SIii\  observed in absorption,
again contrary to observation.

The width of the \SIii \lam 1260 line may result from a combination of factors:
(1)  line-of-sight variations in the expanding shell of unshocked \SIii;
(2) thermal velocity broadening of the shocked \SIii; and (3) a mixture of
unshocked and shocked \SIii, as is required to account for the absorption
profile in the SM star spectrum.   The Gaussian FWHM of 22.7 \AA\ corresponds
to a dispersion of  $\sigma = 2300 \kms$\@.
If we assume that this results entirely from the thermal velocity width
of material heated on passing through the reverse shock, the
reverse shock velocity would be $v_{rs} = \sqrt{16/3}\: \sigma = 5300 \kms$\@.
This is almost twice as great as the reverse shock velocity inferred from the
shocked \SIii\ at far side of the remnant along the line of sight to the SM star and
is likely an overestimate.   Nevertheless, it seems probable that at least some of
the silicon in the core through SN~1006 sampled by the Q1 spectrum has
been shocked.   This is consistent with the assertions made by \citet{long03}  based
both on the clumpiness of the X-ray emission in this region and
on fits to the {\it Chandra} ACIS X-ray spectra that seemed to require
abundances  very different from those expected from shocked interstellar material.
A spectrum with higher signal-to-noise might enable one to sort out how the various
effects contribute to the overall line profile.\footnote{Hamilton \etal (in preparation) are completing
just such a study  for the SM star based on STIS data.}

\begin{figure}[htbp]
\begin{center}
\includegraphics[width=3.5in]{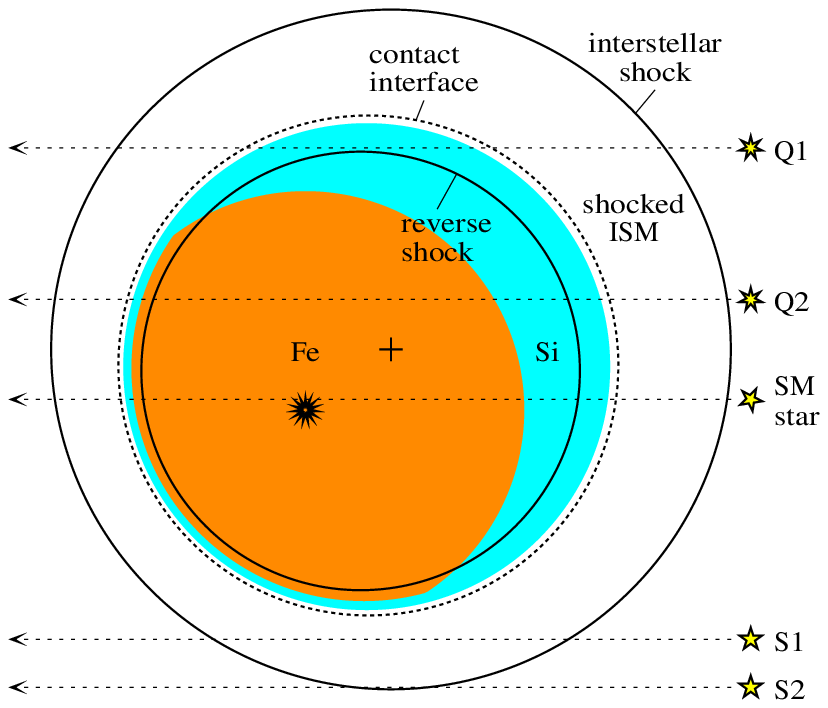}
\end{center}
\caption{
\label{fig8}
Schematic diagram, approximately to scale, of the structure of SN~1006 as
inferred from absorption observations of all five background objects.  Each
line of sight is shown at the correct {\it radial} distance from the projected center of the
outer shell, which is marked with a cross.  Roughly, the diagram represents a
cross-section running from north-northeast
to south-southwest through the center of the image shown in Fig.~\protect\ref{fig1b}.
The explosion center is displaced from the present center of the shell and is
marked by the multi-pointed star.
Iron ejecta (darker shading, rust on the color version) lie in the interior, surrounded by a
mantle of silicon ejecta (lighter shading, blue on the color version).
}
\end{figure}

\subsection{Model Geometry}

Figure~\ref{fig8} illustrates schematically a model for the structure of SN1006,
as inferred from observations of all five background objects.
The radial offsets of the background objects are set  at
their observed radial positions from the geometric center of the remnant,
as listed in Table~1.
Figure~\ref{fig8} shows the remnant as being almost spherical,
whereas HFWCS (their Figure~\ref{fig7}) showed the remnant as being elongated
by about 20\% along the line of sight.
The rounder shape comes from the 20\% greater distance of $2.18 \, {\rm kpc}$
adopted here compared to the $1.8 \, {\rm kpc}$ used by HFWCS.

The center of the original supernova explosion, marked by a cross in Figure~8,
is displaced towards the near side of the remnant, as in HFWCS.
The displacement of the center to the near side
explains how unshocked and low-ionization shocked silicon ejecta
are observed only on the far (redshifted) side, and not on the near (blueshifted)
side of the remnant.
In Figure~\ref{fig8},
the line of sight from the SM star intersects the ejecta
at various velocities measured from observations:
the reverse shock in the iron zone at $-4200 \kms$\ on the near side,
the boundary between the iron and silicon zones
at around $+5600 \kms$\ on the far side,
and the reverse shock in the Si zone
at $+7026 \kms$\ on the far side (HFWCS; see also Hamilton \etal, in preparation).

In order to make the reverse shock nearly tangent to the line of sight to Q1,
as argued above, the explosion center shown in Figure~\ref{fig8} is also displaced
to the southwest.  This geometry also places the SM star closer to the
{\it explosion} center (as distinguished from the present {\it remnant} center)
than Q2.  This provides at least a qualitative explanation for the observation that the blue
edge of the  \feii\  absorption is less blueshifted in Q2 than in the SM star.

Of course the actual geometry is probably more complicated than that shown
schematically in Figure~\ref{fig8}.

\section{Summary and Conclusions}
We have increased to five the total number of sight lines through the SN~1006 remnant
probed by UV absorption spectroscopy.   The new observations, combined with
earlier ones of the SM star, lead to the following conclusions:

(1) The strong, broad Fe and Si absorption lines seen against the SM star
are not an anomaly; similar features are seen in the spectra of the background
quasars Q1 and Q2.  This rules out any possibility that the SM star might have
been the donor star in the pre-SNIa system that gave rise to SN~1006 and that
the broad lines are a local phenomenon in or around the star itself.

(2) Absorption of \feii\ along the line of sight to Q2, which like that to the SM star
passes near the center of the SN~1006 shell, has an equivalent width $\sim 50\%$\ greater
than for the SM star and a similar asymmetric profile.
This suggests that there may be slightly more iron within the SN~1006 shell than
studies of the SM star have indicated; we estimate
$M_{\rm Fe} \simeq 0.06\: M_\sun$ with a $3\:\sigma$\ upper limit of $M_{Fe} < 0.16\: M_\sun$.
This does not change the fundamental conclusion of previous absorption studies
that there is a deficiency of iron within
the SN~1006 shell, compared with the 0.3 to 0.8 $M_{ \sun}$\ that a
Type Ia supernova is expected to produce.
Indeed, demonstration that cold iron is also scarce along a second
line of sight strengthens that basic conclusion.

(3) Strong \SIii\ absorption is seen along the line of sight to Q1,
which passes 8\farcm 3 away from the center (57\% of the shell radius), indicating that much
of the expanding silicon has advanced past this point.  The absorption profile is
consistent with a broad Gaussian with dispersion $\sigma = 2300 \kms$, suggesting that the absorbing
material has been heated by passing through the reverse shock.  Since a single broad
Gaussian seems consistent with the data, most of the material along this line of sight has
probably been shocked, as would be expected if this core sample passes just outside the
present position of the reverse shock.

(4) The absence of \feii\ absorption associated with SN~1006 along the line of sight to
Q1 indicates that the
little if any iron from the SN has advanced this far, and that the ejecta have not
been greatly overturned or mixed.

(5) No unusual absorption lines are seen in the spectra of two background stars
located near the southern limb of the SNR shell.  The silicon and iron ejecta must be confined within
$\sim 85\%$\ of the shell radius, at least at the azimuths of these two stars.

\acknowledgments

We thank Sarah Kate May for her timely reduction of the photometric data that was
necessary for selecting the initial candidate sources for further study, and Gabe Brammer for
assistance with reduction of HST data from the SM star.
P.F.W. and K.S.L. gratefully acknowledge the outstanding support,
typical of the mountain staff at CTIO, during the observations that yielded the
new data reported here.
This work has been made
possible through the financial support  from NASA, through {\it HST} grants NAG 5-8020
to P.F.W. and GO-7349 to A.J.S.H., and {\it Chandra} grants GO0-1120X and GO1-2058A to K.S.L\@.
In addition, P.F.W. acknowledges support from the NSF, through grant
AST-0307613.



\begin{thebibliography}{35}
\expandafter\ifx\csname natexlab\endcsname\relax\def\natexlab#1{#1}\fi

\bibitem[{{Bagnulo} {et~al.}(2003){Bagnulo}, {Jehin}, {Ledoux}, {Cabanac},
  {Melo}, {Gilmozzi}, \& {The ESO Paranal Science Operations Team}}]{bagnulo03}
{Bagnulo}, S., {Jehin}, E., {Ledoux}, C., {Cabanac}, R., {Melo}, C.,
  {Gilmozzi}, R., \& {The ESO Paranal Science Operations Team}. 2003, The
  Messenger, 114, 10

\bibitem[{{Blair} {et~al.}(1996){Blair}, {Long}, \& {Raymond}}]{blair96}
{Blair}, W.~P., {Long}, K.~S., \& {Raymond}, J.~C. 1996, \apj, 468, 871

\bibitem[{{Blair} {et~al.}(2004){Blair}, {Sankrit}, {Torres}, {Chayer},
  {Danforth}, \& {Raymond}}]{blair04}
{Blair}, W.~P., {Sankrit}, R., {Torres}, S.~I., {Chayer}, P., {Danforth},
  C.~W., \& {Raymond}, J.~C. 2004, American Astronomical Society Meeting, 204

\bibitem[{{Bravo} {et~al.}(2004){Bravo}, {Badenes}, \& {Garcia-Senz}}]{bravo04}
{Bravo}, E., {Badenes}, C., \& {Garcia-Senz}, D. 2004, ArXiv Astrophysics
  e-prints, astro-ph/0412155

\bibitem[{{Burleigh} {et~al.}(2000){Burleigh}, {Heber}, {O'Donoghue}, \&
  {Barstow}}]{burleigh00}
{Burleigh}, M.~R., {Heber}, U., {O'Donoghue}, D., \& {Barstow}, M.~A. 2000,
  \aap, 356, 585

\bibitem[{{Cardelli} {et~al.}(1989){Cardelli}, {Clayton}, \&
  {Mathis}}]{cardelli89}
{Cardelli}, J.~A., {Clayton}, G.~C., \& {Mathis}, J.~S. 1989, \apj, 345, 245

\bibitem[{{Drilling} \& {Landolt}(2000)}]{drilling00}
{Drilling}, J.~S. \& {Landolt}, A.~U. 2000, in Allen's Astrophysical
  Quantities, Fourth Edition, ed. A.~N. {Cox} (New York: Springer-Verlag),
  381--396

\bibitem[{{Dyer} {et~al.}(2001){Dyer}, {Reynolds}, {Borkowski}, {Allen}, \&
  {Petre}}]{dyer01}
{Dyer}, K.~K., {Reynolds}, S.~P., {Borkowski}, K.~J., {Allen}, G.~E., \&
  {Petre}, R. 2001, \apj, 551, 439

\bibitem[{{Fesen} {et~al.}(1999){Fesen}, {Gerardy}, {McLin}, \&
  {Hamilton}}]{fesen99}
{Fesen}, R.~A., {Gerardy}, C.~L., {McLin}, K.~M., \& {Hamilton}, A.~J.~S. 1999,
  \apj, 514, 195

\bibitem[{{Fesen} {et~al.}(1989){Fesen}, {Saken}, \& {Hamilton}}]{fesen89}
{Fesen}, R.~A., {Saken}, J.~M., \& {Hamilton}, A.~J.~S. 1989, \apjl, 341, L55

\bibitem[{{Fesen} {et~al.}(1988){Fesen}, {Wu}, {Leventhal}, \&
  {Hamilton}}]{fes88}
{Fesen}, R.~A., {Wu}, C., {Leventhal}, M., \& {Hamilton}, A.~J.~S. 1988, \apj,
  327, 164

\bibitem[{{Ghavamian} {et~al.}(2002){Ghavamian}, {Winkler}, {Raymond}, \&
  {Long}}]{gha02}
{Ghavamian}, P., {Winkler}, P.~F., {Raymond}, J.~C., \& {Long}, K.~S. 2002,
  \apj, 572, 888

\bibitem[{{H{\" o}flich} {et~al.}(1998){H{\" o}flich}, {Wheeler}, \&
  {Thielemann}}]{hoeflich98}
{H{\" o}flich}, P., {Wheeler}, J.~C., \& {Thielemann}, F.~K. 1998, \apj, 495,
  617

\bibitem[{{Hamilton} {et~al.}(1997){Hamilton}, {Fesen}, {Wu}, {Crenshaw}, \&
  {Sarazin}}]{hamilton97}
{Hamilton}, A.~J.~S., {Fesen}, R.~A., {Wu}, C.-C., {Crenshaw}, D.~M., \&
  {Sarazin}, C.~L. 1997, \apj, 481, 838, (HFWCS)

\bibitem[{{Hamuy} {et~al.}(1992){Hamuy}, {Walker}, {Suntzeff}, {Gigoux},
  {Heathcote}, \& {Phillips}}]{hamuy92}
{Hamuy}, M., {Walker}, A.~R., {Suntzeff}, N.~B., {Gigoux}, P., {Heathcote},
  S.~R., \& {Phillips}, M.~M. 1992, \pasp, 104, 533

\bibitem[{{Jenkins} {et~al.}(1998){Jenkins}, {Tripp}, {Fitzpatrick}, {Lindler},
  {Danks}, {Beck}, {Bowers}, {Joseph}, {Kaiser}, {Kimble}, {Kraemer},
  {Robinson}, {Timothy}, {Valenti}, \& {Woodgate}}]{jenkins98}
{Jenkins}, E.~B., {Tripp}, T.~M., {Fitzpatrick}, E.~L., {Lindler}, D., {Danks},
  A.~C., {Beck}, T.~L., {Bowers}, C.~W., {Joseph}, C.~L., {Kaiser}, M.~E.,
  {Kimble}, R.~A., {Kraemer}, S.~B., {Robinson}, R.~D., {Timothy}, J.~G.,
  {Valenti}, J.~A., \& {Woodgate}, B.~E. 1998, \apjl, 492, L147

\bibitem[{{Jenkins} {et~al.}(1976){Jenkins}, {Wallerstein}, \&
  {Silk}}]{jenkins76}
{Jenkins}, E.~B., {Wallerstein}, G., \& {Silk}, J. 1976, \apjs, 32, 681

\bibitem[{{Jenkins} {et~al.}(1984){Jenkins}, {Wallerstein}, \&
  {Silk}}]{jenkins84}
---. 1984, \apj, 278, 649

\bibitem[{{Kasen} {et~al.}(2004){Kasen}, {Nugent}, {Thomas}, \&
  {Wang}}]{kasen04}
{Kasen}, D., {Nugent}, P., {Thomas}, R.~C., \& {Wang}, L. 2004, \apj, 610, 876

\bibitem[{{Long} {et~al.}(2003){Long}, {Reynolds}, {Raymond}, {Winkler},
  {Dyer}, \& {Petre}}]{long03}
{Long}, K.~S., {Reynolds}, S.~P., {Raymond}, J.~C., {Winkler}, P.~F., {Dyer},
  K.~K., \& {Petre}, R. 2003, \apj, 586, 1162

\bibitem[{{Morton}(2003)}]{morton03}
{Morton}, D.~C. 2003, \apjs, 149, 205

\bibitem[{{Schaefer}(1996)}]{schaefer96}
{Schaefer}, B.~E. 1996, \apj, 459, 438

\bibitem[{{Schweizer} \& {Middleditch}(1980)}]{sch80}
{Schweizer}, F. \& {Middleditch}, J. 1980, \apj, 241, 1039

\bibitem[{{Stephenson} \& {Clark}(2002)}]{stephenson02}
{Stephenson}, F.~R. \& {Clark}, D.~H. 2002, {Historical Supernovae and their
  Remnants} (Oxford: Oxford U. Press), 150--174

\bibitem[{{Warren} \& {Hughes}(2004)}]{warren04}
{Warren}, J.~S. \& {Hughes}, J.~P. 2004, \apj, 608, 261

\bibitem[{{Wellstein} {et~al.}(1999){Wellstein}, {Langer}, {Gehren},
  {Burleigh}, \& {Heber}}]{wellstein99}
{Wellstein}, S., {Langer}, N., {Gehren}, T., {Burleigh}, M., \& {Heber}, U.
  1999, in Astronomische Gesellschaft Meeting Abstracts, 3

\bibitem[{{Welsh} {et~al.}(2001){Welsh}, {Sfeir}, {Sallmen}, \&
  {Lallement}}]{welsh01}
{Welsh}, B.~Y., {Sfeir}, D.~M., {Sallmen}, S., \& {Lallement}, R. 2001, \aap,
  372, 516

\bibitem[{{Winkler} {et~al.}(2003){Winkler}, {Gupta}, \& {Long}}]{win03}
{Winkler}, P.~F., {Gupta}, G., \& {Long}, K.~S. 2003, \apj, 531, 829

\bibitem[{{Winkler} \& {Long}(1997{\natexlab{a}})}]{win97b}
{Winkler}, P.~F. \& {Long}, K.~S. 1997{\natexlab{a}}, \apjl, 486, L137

\bibitem[{{Winkler} \& {Long}(1997{\natexlab{b}})}]{win97a}
---. 1997{\natexlab{b}}, \apj, 491, 829

\bibitem[{{Wu} {et~al.}(1993){Wu}, {Crenshaw}, {Fesen}, {Hamilton}, \&
  {Sarazin}}]{wu93}
{Wu}, C., {Crenshaw}, D.~M., {Fesen}, R.~A., {Hamilton}, A.~J.~S., \&
  {Sarazin}, C.~L. 1993, \apj, 416, 247

\bibitem[{{Wu} {et~al.}(1997){Wu}, {Crenshaw}, {Hamilton}, {Fesen},
  {Leventhal}, \& {Sarazin}}]{wu97}
{Wu}, C., {Crenshaw}, D.~M., {Hamilton}, A.~J.~S., {Fesen}, R.~A., {Leventhal},
  M., \& {Sarazin}, C.~L. 1997, \apjl, 477, L53

\bibitem[{{Wu} {et~al.}(1983){Wu}, {Leventhal}, {Sarazin}, \& {Gull}}]{wu83}
{Wu}, C.-C., {Leventhal}, M., {Sarazin}, C.~L., \& {Gull}, T.~R. 1983, \apjl,
  269, L5

\bibitem[{{Zheng} {et~al.}(1997){Zheng}, {Kriss}, {Telfer}, {Grimes}, \&
  {Davidsen}}]{zheng97}
{Zheng}, W., {Kriss}, G.~A., {Telfer}, R.~C., {Grimes}, J.~P., \& {Davidsen},
  A.~F. 1997, \apj, 475, 469

\end{thebibliography}

\clearpage

\clearpage

\begin{deluxetable}{lllrcccrrrrl}
\tabletypesize{\footnotesize}
\rotate
\tablewidth{0pc}
\tablecaption{UV-Bright Sources Behind SN~1006}

\tablehead{
\colhead {}  &
\colhead {} &\colhead {} &
\multicolumn{3}{c}{Distance from Center\tablenotemark{a}} & 
\colhead{} &
\multicolumn{3}{c}{Photometry\tablenotemark{b}} &
\colhead {\phm{l} Spectrum\tablenotemark{c}} &
\colhead {}\\ 

\cline{4-6} \cline{8-10} 

\colhead{Source} & 
\colhead {R.A. (2000.)} &
\colhead {Dec. (2000.)} &
\colhead {\phn \arcmin } & 
\colhead{ \% Radius} &
\colhead{$V_{avg}$} &
\colhead{} &
\colhead {$V$} &
\colhead {$U-B$} &
\colhead {$B-V$} &
\colhead {\phm{l} Exposure} &
\colhead {ID and Comments}
}

\startdata

Star 1502--4207 (S1) &  $15^{\rm h}02^{\rm m}25\fs7 $ & 
$-42\arcdeg 07\arcmin 06\farcs 9$  & 11.9 & 83 & 7420 & &
$16.61 $ & $ -0.16 $ & $0.03$ & 
1200 (2) &   A0V star \\

Star  (SM) &  $15\phm{^{\rm h}}02\phm{^{\rm m}} 53\fs1 $ & 
$-41\phm{\arcdeg} 59\phm{\arcmin} 16. 7$  & 2.8 & 19 & 1720 & &
$16.69 $ & $ -0.94 $ & $-0.19$ & 
4000 (2) &   sdOB star \\

QSO 1503--4155 (Q2) &  $15\phm{^{\rm h}}02\phm{^{\rm m}} 55.2\phm{0} $ & 
$-41\phm{\arcdeg}54\phm{\arcmin} 30.2 $  & 2.0 & 14 & 1280 & &
$19.54 $ & $ -0.80 $ & $ 0.63$ & 
4600 (3) & QSO, $z = 1.026 $ \\

Star 1503--4209 (S2) &  $15\phm{^{\rm h}}03\phm{^{\rm m}}29.5$ & 
$-42\phm{\arcdeg}08\phm{\arcmin} 54.3$  & 13.9 & 92 & 8650 & &
$13.54 $ & $ -0.17 $ & $ -0.01 $ &
600 (1) &  A0V star \\

QSO 1504--4152 (Q1) &  $15\phm{^{\rm h}}03\phm{^{\rm m}}33.9$ & 
$-41\phm{\arcdeg}52\phm{\arcmin} 24.0$ & 8.3 & 57 & 5160 & &
$18.29	$ & $ -0.80 $ & $0.20 $ &
4000 (2) & QSO, $z = 0.337 $ \\

\enddata

\tablenotetext{a}{The center is defined as the center of \ha\ emission, 
$15^{\rm h}02^{\rm m}55\fs 4 $,  $-41\arcdeg 56\arcmin 33\arcsec$\
\citep{win03}. The radius is that of the \ha\ shell 
(Figure 1) at the azimuth of the source.  $V_{avg}$\ represents the average transverse 
velocity (in $\rm{\,km\,s^{-1}}$) of
material just reaching the line of sight, assuming a distance of 2.18 kpc and an age of 995 years.}
\tablenotetext{b}{CTIO 0.9m telescope, 1997 Feb. 10--11.}
\tablenotetext{c} {CTIO 4m Blanco telescope, 1998 June 24--26.  Total exposure time in seconds; 
number of frames in parentheses.}

\end{deluxetable}

\clearpage

\begin{deluxetable}{rccc}
\tablecaption{STIS Observation Log }
\tablehead{\colhead{Source} & 
 \colhead{Obs.~Date} & 
 \colhead{G140L~Exp.} & 
 \colhead{G230L~Exp.} 
\\
\colhead{~} & 
 \colhead{~} & 
 \colhead{(s)} & 
 \colhead{(s)} 
}
\scriptsize
\tablewidth{0pt}\startdata
QSO-1504-4152 (Q1) &  2000-09-09 &  6000 &  2457 \\ 
QSO-1503-4155 (Q2) &  2001-09-06 &  --- &  5457 \\ 
STAR-1503-4209 (S2) &  2000-06-25 &  1526 &  ~600 \\ 
STAR-1502-4207 (S1) &  2000-07-09 &  6000 &  2485 \\ 
\enddata 
\label{tab_obs}
\end{deluxetable}

\end{document}